\documentclass[11pt,journal]{IEEEtran}
\usepackage{graphicx}
\usepackage{color}
\usepackage{placeins}
\usepackage{float}
\usepackage{tabularx,colortbl}

\usepackage{url}                    % for better handling of URL
\usepackage{lscape}                 % allow to use \begin{landscape}, which makes a page in landscape.
\usepackage{subfigure}
\usepackage{mathrsfs}
\usepackage{graphicx}
\usepackage{caption2}
\usepackage{epstopdf}
\usepackage{framed}
\usepackage{xcolor}
\usepackage{citesort}

\usepackage{algorithm}
\usepackage{algorithmic}
\usepackage{multirow}
\usepackage{color}

\usepackage{amssymb}
\usepackage{amsmath}
\usepackage{cite}

\usepackage{subfigure}
\usepackage{citesort}
\usepackage{fancyhdr}
\usepackage{mdwmath}
\usepackage{mdwtab}
\usepackage{flafter}

\begin{document}
\title{Sparse Representation for Wireless Communications: A Compressive Sensing Approach}

\author{
\IEEEauthorblockN{\Large{Zhijin~Qin$^{1}$, Jiancun~Fan$^{2}$, Yuanwei~Liu$^{3}$,
~Yue~Gao$^{3}$, and~Geoffrey Ye Li$^{4}$
 }}

 \IEEEauthorblockA{\small{
 $^{1}$Lancaster University, Lancaster, UK\\
 $^{2}$Xi'an Jiaotong University, Xi'an, China\\
  $^{3}$Queen Mary University of London, London, UK\\
$^{4}$Georgia Institute of Technology, Atlanta, GA, USA\\
%Email: zhijin.qin@lancaster.ac.uk, fanjc@xjtu.edu.cn,\{yuanwei.liu,yue.gao\}@qmul.ac.uk,liye@ece.gatech.edu\\
}}

 }

\maketitle

\begin{abstract}
Sparse representation can efficiently model signals in different applications to facilitate processing. In this article, we will discuss various applications of sparse representation in wireless communications, with focus on the most recent compressive sensing (CS) enabled approaches. With the help of the sparsity property, CS is able to enhance the spectrum efficiency and energy efficiency for the fifth generation (5G) networks and Internet of Things (IoT) networks. This article starts from a comprehensive overview of CS principles and different sparse domains potentially used in 5G and IoT networks. Then recent research progress on applying CS to address the major opportunities and challenges in 5G and IoT networks is introduced, including wideband spectrum sensing in cognitive radio networks, data collection in IoT networks, and channel estimation and feedback in massive MIMO systems. Moreover, other potential applications and research challenges on sparse representation for  5G and IoT networks are identified. This article will provide readers a clear picture of how to exploit the sparsity properties to process wireless signals in different applications.\\


\end{abstract}

\IEEEpeerreviewmaketitle
\section{Introduction}
Sparse representation expresses some signals as a linear combination of a few atoms from a prespecified and over-complete dictionary~\cite{Zibulevsky:2010}. This form of sparse (or compressible) structure arises naturally in many applications~\cite{bruckstein2009sparse}. For example, audio signals are sparse in frequency domain, especially for the sounds representing tones. Image processing can exploit a sparsity property in the discrete cosine  domain, i.e. many discrete cosine transform (DCT) coefficients of images are zero or small enough to be regarded as zero. This type of sparsity property has enabled intensive research on signal and data processing, such as  dimension reduction in data science, wideband sensing in cognitive radio networks (CRNs), data collection in large-scale wireless sensor networks (WSNs), and channel estimation and feedback in massive MIMO.

Traditionally, signal acquisition and transmission adopt the procedure with  sampling and  compression.   As massive connectivity is expected to be supported  in  the fifth generation  (5G) networks and Internet of Things (IoT) networks, the amount of generated data becomes huge. Therefore, signal processing has been confronted with challenges on high sampling rates for data acquisition and large amount of data for storage and transmission, especially in IoT applications with power-constrained sensor nodes. Except for developing more advanced sampling and compression techniques, it is natural to ask whether there is an approach to achieve signal sampling and compression simultaneously.

As an appealing approach  employing sparse representations, compressive sensing (CS) technique~\cite{Candes:2006} has been proposed to reduce data acquisition costs by enabling sub-Nyqusit sampling. Based on the advanced theory~\cite{Candes:robust:2006}, CS has been widely applied in many areas. The key idea of CS is to enable exact signal reconstruction  from far fewer samples than that is required by the Nyquist-Shannon sampling theorem  provided that the signal admits a sparse representation in a certain domain. In CS, compressed samples are acquired via a small set of non-adaptive, linear, and usually randomized measurements, and signal recovery  is usually formulated as an $l_0$-norm minimization problem to find the sparsest solution satisfying the constraints. Since $l_0$-norm minimization is an NP-hard problem, most of the exiting research contributions on CS solve it by either approximating it to a convex $l_1$-norm minimization problem~\cite{Candes:robust:2006} or adopting greedy algorithms, such as orthogonal match pursuit (OMP).

It is often the case that the sparsifying transformation is unknown or difficult to determine. Therefore, projecting a signal to its proper sparse domain is quite essential in many applications that invoke CS. In 5G and IoT networks, the identified sparse domains mainly include frequency domain, spatial domain, wavelet domain, DCT domain, etc. CS can be used to improve spectrum efficiency (SE) and energy efficiency (EE) for these networks. By enabling the unlicensed usage of spectrum,  CRNs exploit spectral opportunities over a wide frequency range  to enhance the network SE. In wideband spectrum sensing, spectral signals naturally exploit a  sparsity property in frequency domain  due to low utilization of spectrum~\cite{Zhijin:TWC:2016,Zhijin:TSP2016}, which enables sub-Nyquist sampling on cognitive devices. Another interesting scenario is a small amount of data collection in large-scale WSNs  with power-constrained sensor nodes, such as smart meter monitoring infrastructure in IoT applications. In particular,  the monitoring readings usually have a sparse representation in  DCT domain due to the temporal and spatial correlations~\cite{Chen:IET2012}. CS can be applied to enhance the EE of WSNs and to extend the lifetime of sensor nodes. Moreover, massive MIMO is a critical technique for 5G networks. In massive MIMO systems, channels corresponding to different antennas are correlated. Furthermore,  a huge number of channel coefficients can be represented by only a few parameters due to a hidden joint sparsity property caused by the shared local scatterers in the radio propagation environment. Therefore, CS can  be potentially used in massive MIMO systems to reduce the overhead for channel estimation and feedback and facilitate precoding~\cite{rao2015compressive}. Even though various applications have different characters, it is worth noting that the signals in different scenarios share a common sparsity property even though the sparse domains can be different, which enables CS to enhance the SE and EE of wireless communications networks.

There have been some interesting surveys on CS~\cite{Candes:SPM:2008} and its applications~\cite{Berger:CM:2010,Sharma:CST:2017,Romero:SPM:2016}. One of the most popular articles on CS~\cite{Candes:SPM:2008} has provided an overview on the theory of CS as  a novel sampling paradigm that goes against the common wisdom in data acquisition. CS-enabled sparse channel estimation has been summarized in~\cite{Berger:CM:2010}. In~\cite{Sharma:CST:2017}, a comprehensive review of the application of CS in CRNs has been provided. A more specific survey on compressive covariance sensing has been presented in~\cite{Romero:SPM:2016} that includes the reconstruction of second-order statistics  even in the absence of prior sparsity information.  These existing surveys serves different purposes. Some cover the basic principles for beginners and others focus on  specific aspects of CS.  Different from the existing literature, our article provides a comprehensive overview of the recent contributions on CS-enabled wireless communications from the perspective of adopting different sparse domain projections.

In this article, we will first introduce the basic principles of CS briefly.  Then we will present the different sparse domains for signals in wireless communications. Subsequently, we will provide CS-enabled frameworks  in various wireless communications scenarios, including wideband spectrum sensing in CRNs, data collection in large-scale WSNs, and channel estimation and feedback for massive MIMO, as they have been  identified to be critical to 5G and IoT networks and share the same spirit by exploiting the sparse domains aforementioned. Within each identified scenario, we start with projecting a signal to a sparse domain, then introduce the CS-enabled framework, and finally illustrate how to  exploit  joint sparsity in the CS-enabled framework. Moreover, the reweighted CS approaches for each scenario will be discussed, where the weights  are constructed by prior information depending on specific application scenarios. The other potential applications and research challenges on applying CS in wireless networks will also be discussed and followed by conclusions.

This article gives readers a clear picture on the research and development of the applications of CS in different scenarios. By identifying the different sparse domains, this article illustrates  the benefits and challenges on applying  CS  in wireless communication networks.

\section{Sparse Representation}
Sparse representation of signals has received extensive attention due to its capacity for efficient signal modelling and related applications. The problem solved by the sparse representation is to search for the most compact representation of a signal in terms of a linear combination of the atoms in an overcomplete dictionary. In the literature, three aspects of research on the sparse representation have been focused:

\begin{enumerate}
  \item Pursuit methods for solving the optimization problem, such as matching pursuit and basis pursuit;
  \item Design of the dictionary, such as the K-SVD method;
  \item Applications of the sparse representation, such as wideband spectrum sensing, channel estimation of massive MIMO, and data collection in WSNs.
\end{enumerate}

%Generally, sparse representation works well in applications where the original signal needs to be reconstructed as accurately as possible.
General sparse representation methods, such as principal component analysis (PCA) and independent component analysis (ICA), aim to obtain a representation that enables sufficient reconstruction. It has been demonstrated that PCA and ICA are able  to deal with signal corruption, such as noise, missing data, and outliers. For  sparse signals without measurement noise,  CS can  recover the sparse signals exactly with random measurements. Furthermore, the random measurements significantly outperform measurements based on PCA and ICA for the sparse signals without corruption~\cite{Chang:2009,Wright2010Sparse,informativesensing:2009}.   In the following, we will focus on the principles of CS and the common sparse domains potentially  used in 5G and IoT scenarios.

\subsection{Principles of Standard Compressive Sensing}
The principles of standard CS, such as to be performed at a single node, can be summarized into the following three parts~\cite{Candes:2006}:
\subsubsection{Sparse Representation}
Generally speaking, sparse signals contain much less information than their ambient dimension suggests. Sparsity of a signal is defined as the number of non-zero elements in the signal under a certain domain. Let $\textbf{f}$ be an $N$-dimensional signal of interest, which is sparse over  the orthonormal transformation basis matrix $\mathbf{\Psi}  \in {\mathbb{R}^{N \times N}}$, and $\textbf{s}$ be the sparse representation of $\textbf{f}$ over the basis $\mathbf{\Psi}$. Then $\mathbf{f}$ can be given by
\begin{equation}
   \textbf{f} = \mathbf{\Psi} \textbf{s}.
      \label{equ1:sparse}
\end{equation}
Apparently, $\textbf{f}$ can be the time or space domain representation of a signal, and $\textbf{s}$ is the equivalent representation of $\textbf{f}$ in  the $\mathbf{\Psi}$ domain. For example, if $\mathbf{\Psi}$ is the inverse Fourier transform (FT) matrix, then $\textbf{s}$ can be regarded as the frequency domain representation of the time domain signal, $\textbf{f}$. Signal $\textbf{f}$ is said to be $K$-sparse in the $\mathbf{\Psi}$ domain if there are only $K \left(K  \ll  N\right)$ out of the $N$ coefficients in $\textbf{s}$ that are non-zero. If a signal is able to be sparsely represented in a certain domain, the CS technique can be invoked to take only a few linear and non-adaptive measurements.

\subsubsection{Projection}
When the original signal $\mathbf{f}$ arrives at the receiver, it is processed by the measurement matrix $\mathbf{\Phi} \in \mathbb{R}^{P \times N}$ with $ {P < N} $, to get the compressed version of the signal, that is,
     \begin{equation}
         \textbf{x}= \mathbf{ \Phi} \textbf{f} =\mathbf{ \Phi \Psi} \textbf{s}{\rm{ = }}\mathbf{\Theta} \mathbf{s},
         \label{equ2:projection}
     \end{equation}
where $\mathbf{\Theta}  = \mathbf{ \Phi\Psi} $ is an $P \times N$  matrix, called the sensing matrix. As $\mathbf{\Phi}$ is independent of signal $\textbf{f}$, the projection process is non-adaptive.

\begin{figure*}[!t]
\centering
\includegraphics[width=5.8in]{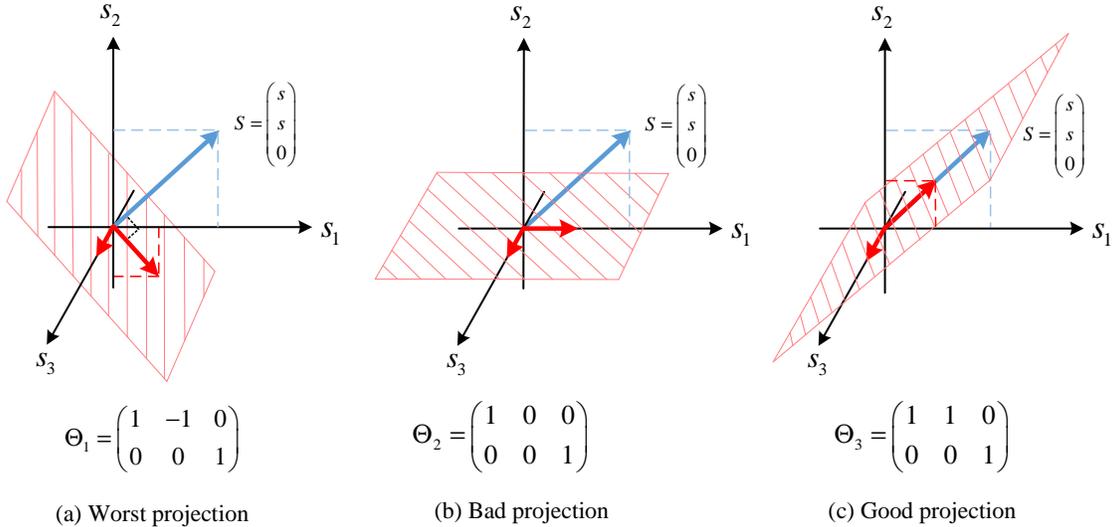}
\caption{Projection of a sparse signal with one non-zero component with different sensing matrices.}
\label{projection}
\end{figure*}

Fig.~\ref{projection} illustrates how the different sensing matrices $\mathbf{\Theta}$ influence the projection of a signal from high dimension to its space, i.e., mapping $\textbf{s}\in \mathbb{R}^3$ to  $\textbf{x} \in \mathbb{R}^2$. As shown in Fig.~\ref{projection}, $\textbf{s} = \left( {\begin{array}{*{20}{c}}
s&s&0
\end{array}} \right)$ is a three-dimensional signal. When $\mathbf{s}$ is mapped into a two-dimensional space by taking $\mathbf{\Theta _1} = \left( {\begin{array}{*{20}{c}}
1&{ - 1}&0\\
0&0&1
\end{array}} \right)$ as the sensing matrix, the original signal $\mathbf{s}$ cannot be recorded based on the projection under $\mathbf{\Theta_1}$. This is because that the plane spanned  by the two row vectors of $\mathbf{\Theta_1}$ is orthogonal to signal $\mathbf{s}$ as shown in Fig.~\ref{projection}(a). Therefore, $\mathbf{\Theta_1}$ corresponds to the worst projection. As shown in Fig.~\ref{projection}(b), we can also observe that the projection by taking $\mathbf{\Theta _2} = \left( {\begin{array}{*{20}{c}}
1&0&0\\
0&0&1
\end{array}} \right)$  is not a good one. It is noted that the plane spanned  by the two row vectors of $\mathbf{\Theta _2}$ can only contain part of information of the sparse signal $\mathbf{s}$, and the  sparse component in the direction of $\mathbf{s_2}$ is missed when the signal $\mathbf{s}$ is projected into the two-dimensional space. When the sensing matrix is set to $\mathbf{\Theta _3} = \left( {\begin{array}{*{20}{c}}
1&1&0\\
0&0&1
\end{array}} \right)$, as shown in Fig.~\ref{projection}(c), the signal $\mathbf{s}$ can be fully recorded as it falls into the plane spanned  by the two row vectors of $\mathbf{\Theta_3}$. Therefore, $\mathbf{\Theta_3}$ results in a good projection and $\mathbf{s}$ can be exactly recovered by its projection $\mathbf{x}$ in the two-dimensional space. Then it is natural to ask what type of  projection is good enough to guarantee the exact signal recovery?

The key of CS theory is to find out a stable basis $\mathbf{\Psi}$ or measurement matrix $\mathbf{\Phi}$ to achieve exact recovery of the  signal with length $N$ from $P$ measurements. It seems an undetermined problem as $P<N$. However, it has been proved in~\cite{Candes:robust:2006} that exact recovery can be guaranteed under the following conditions:

\begin{itemize}
  \item Restricted isometry property (RIP): Measurement matrix $\mathbf{\Phi}$ has the RIP of order $K$ if
  \begin{align}
  1 - {\delta _K} \le \frac{{\left\| {\mathbf{\Phi} \mathbf{f}} \right\|_{\ell_2}^2}}{{\left\| \mathbf{f} \right\|_{\ell_2}^2}} \le 1 + {\delta _K}
  \end{align}
 holds for all $K$-sparse signal $\mathbf{f}$, where ${\delta _K}$ is the restricted isometry constant of a matrix $\mathbf{\Phi}$.
  \item Incoherence property: Incoherence property requires that the rows of measurement matrix $\mathbf{\Phi}$ cannot sparsely represent the columns of  the sparsifying matrix $\mathbf{\Psi}$ and vice verse. More specifically, a good measurement will pick up a little bit information of each component in $ \mathbf{s}$ based on the condition that $\mathbf{\Phi}$ is incoherent with $\mathbf{\Psi}$. As a result, the extracted information can be maximized by using the minimal number of measurements.
\end{itemize}

It has been pointed out  that verifying both the RIP condition and incoherence property is computationally complicated  but they could be achieved with a high probability simply by selecting $\mathbf{\Phi}$ as a random matrix. The common random matrices include Gaussian matrix, Bernoulli matrix, or almost all others matrices with independent and identically distributed (i.i.d.) entries. Besides, with the properties of the matrix with i.i.d. entries $\mathbf{\Phi}$, the matrix $\mathbf{\Theta}=\mathbf{\Phi \Psi}$ is also random i.i.d., regardless of the choice of $\mathbf{\Psi}$. Therefore, the random matrices are universal as they are random enough to be incoherent with any fixed basis. It has been demonstrated that random measurements can universally capture the information relevant for many compressive signal processing applications without any prior knowledge of either the signal class and its sparse domain or the ultimate signal processing task.
%Another commonly used combination is to take the spike basis for the measurement matrix $\mathbf{\Phi}$ and the Fourier basis for the sparsifying matrix $\mathbf{\Psi}$. This type of measurement matrix is easy to deploy at the cost of performance degradation on signal recovery.

\begin{figure}[!t]
\centering
\includegraphics[width=2.8in]{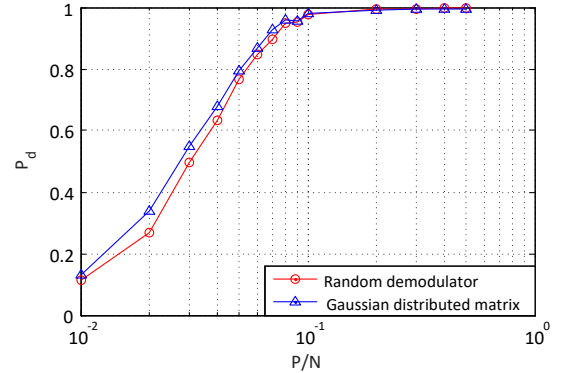}
\caption{Detection probability versus compression ratio with different measurement matrices. In this case, the signal is one-sparse and the simulation iteration is 1000.}
\label{different_matrix}
\end{figure}

Moreover, for Gaussian matrices the number of measurements required to guarantee the exact signal recovery is almost minimal. However, random matrices inherently have two major drawbacks in practical applications: huge memory buffering for storage of matrix elements, and high computational complexity due to their completely unstructured nature~\cite{candes2007sparsity}. Compared to the standard CS that limits its scope to standard discrete-to-discrete measurement architectures using random measurement matrices and signal models based on standard sparsity,  more structured sensing architectures, named structured CS, have been proposed to implement CS on feasible acquisition hardware. So far, many efforts have been put on the design of structured CS matrices, i.e., random demodulator~\cite{Tropp:2010}, to make CS implementable with expense of performance degradation. Particularly, the main principle of random demodulator is to multiply the input signal with a high-rate pseudonoise sequence, which spreads the signal across the entire spectrum. Then a low-pass anti-aliasing filter is applied and the signal is captured by sampling it at a relatively low rate. With the additional digital processing to reduce the burden on the analog hardware, random demodulator bypasses the need for a high-rate analogue-to-digital converter (ADC)~\cite{Tropp:2010}. A comparison of Gaussian sampling matrix  and random demodulator is provided in Fig.~\ref{different_matrix} in terms of detection probability with different compression ratios $P/N$. From the figure, the Gaussian sampling matrix  performs better than the random demodulator.

\subsubsection{Signal Reconstruction}
After the compressed measurements are collected, the original signal should be reconstructed. Since most of the basis coefficients in $\textbf{s}$ are negligible, the original signal can be reconstructed by finding out the minimal set of coefficients that matches the set of compressed measurements $\textbf{x}$, that is, by solving
\begin{equation}
      \mathbf{\hat s}= \arg\min_{\mathbf{s}} {\left\| \mathbf{s} \right\|_{\ell_p}} \,\,\, \text{subject}\,\,\text{to}\,\, \mathbf{\Theta} { \mathbf{s} }= \textbf{x},
      \label{equ3:recovery}
\end{equation}
where ${\left\|  \cdot  \right\|_{{\ell_p}}}$ is the $\ell_p$-norm  and $p=0$ corresponds to  counting the number of non-zero elements in $\mathbf{s}$. However,  the reconstruction problem in~\eqref{equ3:recovery} is both numerically unstable and NP-hard~\cite{Candes:2006} when $\ell_0$-norm is used.

So far, there are mainly two types of relaxations to problem~(\ref{equ3:recovery}) to find a sparse solution. The first type is convex relaxation, where ${\ell_1}$-norm is used to substitute $\ell_0$-norm in~(\ref{equ3:recovery}). Then~(\ref{equ3:recovery}) can be solved by standard convex solvers, e.g., cvx. It has been proved that $\ell_1$ norm results in the same solution as $\ell_0$ norm when RIP is satisfied with the constant ${\delta _{2k}} < \sqrt 2  - 1$~\cite{CANDES2008589}. Another type of solution is to use a greedy algorithm, such as OMP~\cite{Tropp:2007:TIT}, to find a local optimum in each iteration. In comparison with the convex relaxation, the greedy algorithm usually requires lower computational complexity and time cost, which makes it more practical for wireless communication systems. Furthermore, the recent result has shown that the recovery accuracy achieved by some greedy algorithms is comparable to the convex relaxation but requiring much lower computational cost~\cite{Choi:Tutorials:2017}.

\subsection{Reweighted Compressive Sensing}\label{WCS}
As aforementioned, $\ell_1$-norm is a good approximation for the NP-hard $\ell_0$-norm problem when RIP holds. However,   the large coefficients are penalized more heavily than the small ones in $\ell_1$-norm minimization, which leads to performance degradation on signal recovery. To balance the penalty on the large and the small coefficients, reweighted CS is introduced by providing different penalties on those large and small coefficients. A reweighted $\ell_1$-norm minimization framework~\cite{Candes:2006} has been developed to enhance the signal recovery performance with fewer compressed measurements by solving
\begin{align}
{\bf{\hat s}} = \arg \mathop {\min }\limits_{\textbf{s}} {\left\| {\bf{W}{\textbf{s}}} \right\|_{\ell_1}}\;\;\; {\rm{subject}}\;\;{\rm{to}}\;\;{\bf{\Theta \textbf{s}}} = {\textbf{x}},
\end{align}
where $\mathbf{W}$ is a diagonal matrix with ${w_1}, \ldots ,{w_n}$ on the diagonal and zeros elsewhere.

Moreover, $\ell_p$-norm, e.g. $0 < p < 1$, is utilized to lower the computational complexity of signal recovery process caused by the $\ell_1$-norm optimization problem. Iterative reweighted least-square (IRLS) based CS approach has been proposed in~\cite{rao:1999} to solve~\eqref{equ3:recovery} in a non-convex approach as
\begin{align}\label{IRLS}
{\bf{\hat s}} = \arg \mathop {\min }\limits_{\textbf{s}} \sum\limits_{i = 1}^N {{w_i}{s_i}}\;\;\; {\text{subject}}\;\; {\text{to}}\;\; {\bf{\Theta \textbf{s}}} = {\textbf{x}},
\end{align}
where ${w_i} = {\left| {s_i^{\left( {l - 1} \right)}} \right|^{p - 2}}$ is computed based on the result of the last iteration, $s_i^{\left( {l - 1} \right)}$.

It is worth noting that~\eqref{equ3:recovery} becomes non-convex when $p<1$. The existing algorithms cannot guarantee to reach a global optimum and may only produce local minima. However, it has been proved~\cite{Chartrand:2007,Chartrand_restricted:2008} that under some circumstances the reconstruction in~\eqref{equ3:recovery} will reach a \emph{unique} and \emph{global} minimizer~\cite{Chartrand_Yin:2008}, which is exactly $\bf{\hat s}=\textbf{s}$. Therefore, we can still exactly recover the  signal in practice.

\subsection{Distributed Compressive Sensing}
The distributed compressive sensing (DCS)~\cite{Wakin:DCS:2005} is an extension of the standard one by considering networks with $M$ nodes. At the $m$-th node, measurement $x_m$ can be given by
\begin{align}\label{DCS}
{x_m} = {\Theta _m}{s_m},\;\; \forall m \in \mathcal{M},
\end{align}
where $\mathcal{M}$ is the the set of nodes in the network. As stated in~\eqref{equ2:projection}, ${\Theta _m}$ is the sensing matrix deployed at the $m$-th node, and  $s_m$ is a sparse signal of interest. DCS becomes a standard CS when $M=1$.

In the applications of standard CS, the signal received at the same node has its sparsity property due to its intra-correlation. While for the networks with multiple nodes, signals received at different nodes exhibit  strong inter-correlation. The intra-correlation and inter-correlation of signals from the multiple nodes lead to a joint sparsity property. The joint sparsity level  is usually smaller than the aggregate over the individual signal's sparsity level. As a result, the number of compressed measurements required for exact recovery in DCS can be reduced significantly compared to the case performing standard CS at each single node independently.

In DCS, there are two closely related concepts: distributed networks and distributed CS solvers. The distributed networks refer to networks that different  nodes perform data acquisition in a distributed way and the standard CS can be applied at each node individually to perform signal recovery. While for DCS solver as proposed in~\cite{Wakin:DCS:2005}, the data acquisition process requires  no collaboration among  sensors and the signal recovery process is performed at several computational nodes, which can be distributed in a network or locally placed within a multiple core processor. Generally,  it is of interest to minimize both computation cost and communication overhead in DCS. The most popular application scenario of DCS  is that all signals share the common sparse support but with different non-zero coefficients. %A comprehensive summary for different types of signal models for CS can be found in~\cite{sundman2013methods}.

%we are typically interested in improving sinal recovery accuracy or reducing number of measurements at nodes by assuming that data collected at different nodes are highly correlated.

\subsection{Common Sparse Domains for CS-enabled 5G and IoT Networks}
CS-enabled sub-Nyquist sampling is possible only if the signal is sparse in a certain domain. The common sparse domains utilized in CS-enabled 5G and IoT networks include frequency domain, wavelet domain, discrete cosine domain, angular domain, to name a few.

\begin{enumerate}
  \item Frequency domain: due to low spectrum utilization, the wideband spectrum signal shows a sparse property when it is converted into the frequency domain.
    \item Discrete cosine domain: due to  temporal correlation, signals in some applications, such as environment information monitoring, show a sparse property in the discrete cosine domain as the readings normally do not change too much within a short period.
  \item Spatial domain: as the number of paths as well as the angle-of-arrival are much smaller than the number of antennas in massive MIMO systems, the channel conditions can be represented by a limited number of parameters. In this case, the spatial domain turns into the angular domain.
\end{enumerate}

\begin{figure}[!t]
\centering
\includegraphics[width=2.0in]{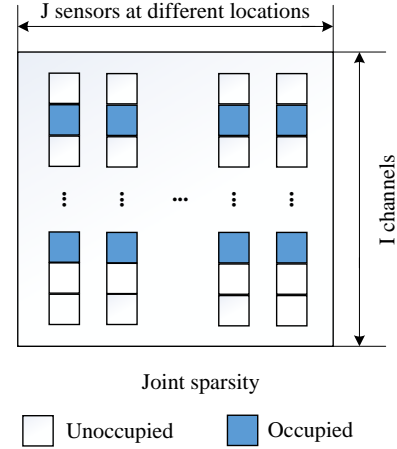}
\caption{Spatial-frequency correlation.}
\label{joint_sparsity}
\end{figure}
For multi-node cases, due to the spatical correlation, the joint sparsity is exploited to apply DCS in  spatial-x domains, where `x' can be any of the aforementioned domains. Here, we give some examples on how the joint sparsity is utilized in different scenarios in 5G and IoT networks:

\begin{enumerate}
  \item Spatial-frequency domain: An illustration of the DCS-enabled cooperative CRN is given in Fig.~\ref{joint_sparsity}, where the joint sparsity in the spatial-frequency domain is utilized. Specifically, each column represents the signal received at each location, which is sparse in frequency domain as only a few channels are occupied. At different locations, the same frequency bands may be occupied, but the signal powers for each frequency band are various due to fading and shadowing. Therefore, different columns of the matrix share  the common sparse support though each node operates without cooperation. With the DCS framework, each node performs sub-Nyquist sampling individually first  and then the original signals can be recovered simultaneously. More details on this issue will be discussed in Section~\ref{III}.
   \item Spatial-temporal domain: In WSNs, sensor nodes are deployed to periodically monitor data and to send the compressed data to the sink. Then the sink is responsible for recovering the original reading by CS algorithms as the readings across all sensor nodes exhibit both spatial and temporal correlations, as we can see from the discussion in Section~\ref{IV}.
   \item Angular-time domain: Massive MIMO channels between some users and the massive base station (BS) antennas appear the spatial common sparsity in both  the time domain and the angular domain, as we will discuss in detail in Section~\ref{V}.

\end{enumerate}

Different sparse domains and their applications in 5G and IoT networks are summarized in Table~I, where `WT' is short for wavelet transform. In addition to the listed sparse domains and their applications, it is worth noting that the core of applying CS is to identify how to exploit the sparse property in 5G and IoT networks. In the following, we discuss three major applications of CS in 5G and IoT networks.

\begin{table*}[!t]
\label{Sparse_domain_table}
\centering
\caption{Common sparse domains and their applications in 5G and IoT networks.}
\begin{tabular}{| p{0.15\linewidth}| p{0.1\linewidth}|p{0.2\linewidth}|p{0.25\linewidth}|p{0.16\linewidth}|}
\hline
\textbf{Sparse domain  }                        & \textbf{Sparsifying transform}             & \textbf{Applications}                             & \textbf{Why sparse?}                                       & \textbf{Sparsity property} \\ \hline
Frequency domain                       & FT                & Wideband spectrum sensing                & Low spectrum utilization in practice                         & Single sparsity   \\ \hline
Spatial domain                         & -                & Channel estimation  in massive MIMO                     & Number of paths is much fewer than the number of antennas       & Single sparsity   \\ \hline
Discrete cosine domain                 & DCT         & Sensor data gathering                    & Temporal correlation                             & Single sparsity   \\ \hline
%Angular domain                           & FT                & Channel estimation in massive MIMO                        &       Number of angle-of-arrival is much less than the number of antennas.                                             & Single sparsity   \\ \hline
Wavelet domain                         & WT               & Sensor data gathering                    & Temporal correlation                             & Single sparsity   \\ \hline
Spatial-frequency domain               & FT                & Cooperative wideband spectrum sensing    & Spatial correlation and low spectrum utilization & Joint sparsity    \\ \hline
Spatial-discrete cosine/wavelet domain & DCT/WT & Active node detection/data gathering & Spatial and temporal correlation                 & Joint sparsity    \\ \hline
Angular-time domain & - & Channel estimation in massive MIMO & Number of paths and degrees of arrival are much fewer than the number of antennas               & Joint sparsity    \\ \hline
\end{tabular}
\end{table*}

\section{Compressive Sensing enabled Cognitive Radio Networks}\label{III}
In this section, we introduce the applications of CS in CRNs.
%\subsection{Cognitive Radio Network}
Radio frequency (RF) spectrum is a valuable but tightly regulated resource due to its unique and important role in wireless communications. The demand for RF spectrum is increasing due to a rapidly expanding market of multimedia wireless services while the usable spectrum is becoming scarce due to the current rigid spectrum allocation policies. However, according to the reports from the Federal Communications Commission (FCC) and the Office of Communications (Ofcom), localized temporal and geographic spectrum utilization is extremely low and unbalanced in reality. Cognitive radio (CR) has become a promising solution to solve the spectrum scarcity problem, by allowing unlicensed secondary users (SUs) to opportunistically access a licensed band when the licensed primary user (PU) is absent. In order to avoid any harmful interference to the PUs, SUs in CRNs should be aware of the spectrum occupancy over the spectrum of interest. Spectrum sensing, which  detects the spectrum holes, the frequency bands that are not utilized by PUs, is one of the most challenging tasks in CR. As the radio environment changes over time and space, an efficient spectrum sensing technique should be capable of tracking these fast changes~\cite{Ying-chang:2008}. A good approach for detecting PUs is to adopting the traditional narrowband sensing algorithms, which include energy detection, matched-filtering, and cyclostationary feature detection. Here, the term narrowband implies that the frequency range is sufficiently narrow such that the channel frequency response can be considered as flat. In other words, the bandwidth of interest is less than the coherence bandwidth of the channel~\cite{sun_wideband:2013}.

While the present spectrum sensing algorithms have focused on exploiting spectral opportunities over a narrow frequency band, future CRNs will eventually be required to exploit spectral opportunities over a wide frequency range from hundreds of megahertz (MHz) to several gigahertz (GHz), in order to improve spectrum efficiency and achieve higher opportunistic throughput. As driven by the Nyquist-Shannon sampling theory, a simple approach is to acquire the wideband signal directly by a high-speed  ADC, which is particularly challenging or even unaffordable, especially for energy-constrained devices, such as smart phones or even battery-free devices. Therefore, revolutionary wideband spectrum sensing techniques become more than desired to release the burden on high-speed ADCs.

\subsection{Standard Compressive Spectrum Sensing}
Recent development on CS theory inspires sub-Nyquist sampling, by utilizing the sparse nature of signals~\cite{Candes:2006}. Driven by the inborn nature of the sparsity property of signals in wireless communications, e.g., the sparse utilization of spectrum, CS theory is capable of enabling sub-Nyquist sampling for wideband spectrum sensing.

\subsubsection{Energy Detection based Compressive Spectrum Sensing}
\begin{figure*}[!t]
\centering
\includegraphics[width=5.3in]{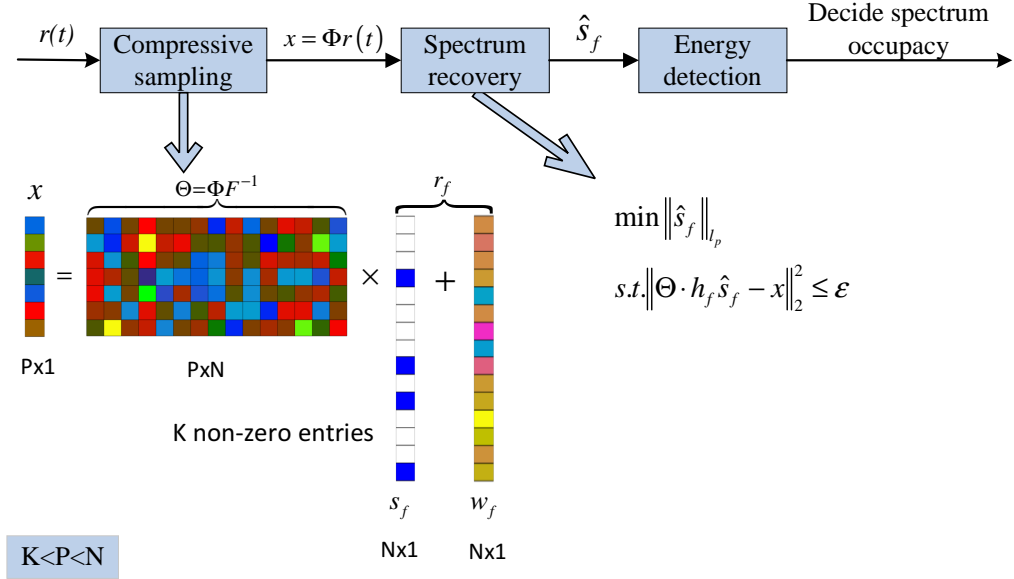}
\caption{Framework of compressive spectrum sensing with energy detection.}
\label{CSS:framework}
\end{figure*}

CS theory has been applied to wideband spectrum sensing in~\cite{zhitian:2007}, where sub-Nyquist sampling is achieved without loss of any information.  A general framework for compressive spectrum sensing with the energy detection method is summarized as shown in Fig.~\ref{CSS:framework}, where the analog signal at the receiver, $r\left( t \right)$ has a sparse representation $s_f$ in the frequency domain. The received signals are then sampled at a sub-Nyquist rate. Due to  low  spectrum utilization, $s_f$ can be recovered from the under-sampled measurements. Then the energy of each channel can be calculated, and therefore the spectrum occupancy can be determined.

Lately, it has been identified in~\cite{Davenport:2012} that the CS-enabled system is somewhat sensitive to noise, exhibiting a 3 dB SNR loss per octave of subsampling, which parallels the classic noise-folding phenomenon. In order to improve  robustness to noise, a denoised compressive spectrum sensing algorithm has been proposed in~\cite{Zhijin:TSP2016}. The sparsity level is required in advance in order to determine the lower sampling rate at SUs locally without loss of any information. However, sparsity level is dependent on spectrum occupancy, which is usually unavailable in dynamic CR networks. In order to solve this problem, a have proposed a two-step CS scheme has been proposed in~\cite{yue_sparsity:2012}  to minimize the sampling rates when the sparsity level is changing. In this approach, the actual sparsity level is estimated first and the number of compressed measurements to be collected is then adjusted before sampling. However, this algorithm introduces extra computational complexity by performing the sparsity level estimation. In order to avoid sparsity level estimation, Sun~\emph{et al.}~\cite{Hongjian:2012} have proposed to adjust the number of compressed measurements adaptively by acquiring compressed measurements step by step in continuous sensing slots. However, this iterative process incurs higher computational complexities at the SU as~\eqref{equ3:recovery} has to be solved several times until the exact signal recovery is achieved. A low-complexity compressive spectrum sensing algorithm has been proposed in~\cite{Zhijin:TWC:2016} by alleviating the iterative process of signal recovery. More specifically, geolocation data can provide a rough estimation of the sparsity level to minimize the sampling rates. Subsequently, data from geolocation database is utilized as the prior information for signal recovery. By doing so, signal recovery performance is improved with significant reduction on the computational complexity and minimal number of measurements.

\subsubsection{Compressive Power Spectral Density Estimation}
Different from the aforementioned approaches that concentrate on spectral estimation with perfect reconstruction of the original signals, compressive power spectral density (PSD) estimation provides another approach for spectrum detection without requiring complete recovery of the original signals. Compressive PSD estimation has been widely applied as the original signals are not actually required in many signal processing applications. For wideband spectrum sensing applications with spectrally sparse signals, only the PSD, or equivalently the autocorrelation function, needs to be recovered as only the spectrum occupancy status is required for each channel.

Polo~\emph{et~al.}~\cite{Polo:2009} have proposed to reconstruct the autocorrelation of the compressed signal to provide an estimate of the signal spectrum by utilizing the sparsity property of the edge spectrum, in which the CS is performed directly on the wide-band analog signal. Nevertheless, the compressive measurements are assumed to be wide-sense stationary in~\cite{Polo:2009}, which is not true for some compressive measurement matrices. Subsequently, Lexa~\emph{et~al.}~\cite{Lexa:ICASSP:2011} have proposed a multicoset sampling based power spectrum estimation method, by exploiting the fact that a wide-sense stationary signal corresponds to a diagonal covariance matrix of the frequency domain representation of the signal. Additionally, Leus~\emph{et~al.}~\cite{Leus:SPL:2011} have solved  the power spectrum blind sampling problem based on a periodic sampling procedure and have further proposed a simple least-square (LS) reconstruction method for power spectrum recovery.

\subsubsection{Beyond Sparsity}
For spectrum blind sampling, the goal is to perfectly reconstruct the spectrum and sub-Nyquist rate sampling is only possible if the spectrum is sparse. However, sub-Nyquist rate sampling can be achieved in~\cite{Leus:SPL:2011} without making any constraints on the power spectrum, but the LS reconstruction requires some rank conditions to be satisfied. Leus~\emph{et~al.}~\cite{Ariananda:2012} have further proposed an efficient power spectrum reconstruction and a novel multicoset sampling implementation by exploiting the spectral correlation properties, without requiring any sparsity constraints on the power spectrum. More recently, Cohen and Eldar~\cite{Cohen:2014} have developed a compressive power spectrum estimation framework for both sparse and non-sparse signals as well as blind and non-blind detection in the sparse case. For each one of those scenarios, the minimal sampling rate allowing perfect reconstruction of the signal's power spectrum is derived in a noise-free environment.

\subsection{Cooperative Spectrum Sensing with Joint Sparsity}
In spectrum sensing, the performance is degraded by noise uncertainty, deep fading, shadowing and hidden nodes. Cooperative spectrum sensing (CSS) has been proposed to improve sensing performance by exploiting the collaboration among all the participating nodes. In CRNs, a CSS network constructs a multi-node network. As aforementioned, joint sparsity property and low-rank property can be utilized to recover the original signals simultaneously with fewer  measurements and the DCS is utilized as it fits the CSS model perfectly. In the existing literature, cooperative compressive spectrum sensing mainly include two categories: i) centralized approaches; ii) decentralized approaches.

A centralized approach  involves a fusion center (FC) performing signal recovery by utilizing the compressed measurements contributed by the spatially distributed SUs. In~\cite{Zhijin:TSP2016}, a robust wideband spectrum sensing algorithm has been proposed for centralized CSS. Specifically, each SU  senses a segment of the spectrum at sub-Nyquist rate to reduce the sensing burden. With the collected compressive measurements, CS recovery algorithms are performed at the FC in order to recover the original signals by exploiting the sparse nature of spectrum occupancy. It is worth noting that the sparse property of signals received at the SUs can be transformed into the low-rank property of the matrix constructed at the FC. In the case of CSS with sub-Nyquist sampling, the measurements collected by the participating SUs are sent to the FC, where the joint sparsity or low-rank property is exploited to recover the complete matrix.

A typical decentralized approach has been proposed in~\cite{Tian:2011}, in which the decentralized consensus optimization algorithm can attain high sensing performance at a reasonable computational cost and power overhead by utilizing the joint sparsity property. Different from the centralized approach, signal recovery or matrix completion is performed at each individual node in the decentralized approaches.  Compared with the centralized approaches, the decentralized ones are more robust as it adopts a FC-free network structure. Another advantage of the decentralized approaches is that they allow the recovery of individual sparse components at each node as well as the common sparse components shared by all participating nodes.

Furthermore, the privacy and security issues in CSS have been investigated in~\cite{Zhijin:TSP:2017} by exploiting joint sparsity in the frequency and spatial domains. In~\cite{Zhijin:TSP:2017},  measurements corrupted by malicious users are removed during the signal recovery process at the FC  so that  the  recovery accuracy and security of the considered networks can be improved.

\subsection{Compressive Spectrum Sensing with Prior Information}
In conventional compressive spectrum sensing, only the sparsity property is utilized. Certain prior information is available in some scenarios and can be exploited to improve performance of wideband spectrum sensing in CRNs. For example, in the case of spectrum sensing over TV white space (TVWS), where the PUs are TV signals and the transmitted waveforms are determined by the standard, this prior information together with the specifications dictated by the spectrum regulatory bodies, i.e., carrier frequencies, bandwidths, can be also utilized to enhance the signal recovery performance. Thus, it is reasonable to assume that the PSD of the individual transmission is known up to a scaling factor.

As discussed in~\ref{WCS}, reweighted CS normally introduces  weights  to provide different penalties on large and small coefficients, which naturally inspires the application of reweighted CS in wideband spectrum sensing with available prior information. In~\cite{Liu:HASP:2012}, the whole spectrum is divided into different segments as the bounds between different types of primary radios are known in advance. Within each segment, an iteratively reweighted $\ell_1$/$\ell_2$ formulation has been proposed to recover the original signals. In~\cite{Zhijin:TWC:2016}, a low-complexity wideband spectrum sensing algorithm for the TVWS spectrum has been proposed to improve the signal recovery performance, in which the weights are constructed by utilizing the prior information from the geolocation database. For example, in the TVWS spectrum, there are 40 TV channels and each channel spans over 8 MHz that can be either occupied or not. Hence, the TV signals show a group sparsity property in the frequency domain as the non-zero coefficients show up in clusters. A more efficient approach has been developed in~\cite{Eldar:Block_sparse:2010} by utilizing such group sparsity property. Moreover, the signals in wideband spectrum sensing have the following two characteristics: i) the input signals are stationary so that their covariance matrices are redundant; ii) most information in practical signals is concentrated on the first few lags of the autocorrelation. Inspired by these characteristics, a spectral prior information assisted structured covariance estimation algorithm has been proposed in~\cite{Romero:TSP2013} with low-computational complexity, which especially fits in application on low-end devices.

\subsection{Potential Research}
We have reviewed some research results in CS-enabled CRNs. There are still many open research issues in the area, especially when practical constraints are considered. In this section, we will introduce a couple of important ones.

\subsubsection{Performance Limitations under Practical Constraints}
Although there exist many research contributions  in the field of compressive spectrum sensing, most of them have assumed some ideal operating conditions. In practice, there may exist various imperfection, such as noise uncertainty, channel uncertainty, dynamic spectrum occupancy, and transceiver hardware imperfection~\cite{Sharma:CST:2017}. For example, the centralized compressive spectrum sensing normally considers ideal reporting channels, which is not the case in practice. This imperfection may lead to significant performance degradation in practice. Another example comes from the measurement matrix design. As shown in Fig.~\ref{different_matrix}, the Gaussian distributed matrix achieves better performance but with a higher implementation cost. Even through some structured measurement matrices, such as random demodulator, with a lower cost and acceptable recovery performance degradation, have been proposed to enable the implementation of CS as a replacement of high-speed ADCs, the nonlinear recovery process limits its implementation. Therefore, it is a big challenge to further investigate compressive spectrum sensing  in the presence of practical imperfection and to develop a common framework to combat their aggregate effects in CS-enabled CRNs.

\subsubsection{Generalized Platform for Compressive Spectrum Sensing}
The existing hardware implementation of sub-Nyquits sampling system follows the procedure that the theoretic algorithm is specifically designed for the current available hardware devices. However, it is very difficult or sometimes even impossible to extend the current hardware architectures to implement other existing compressive spectrum sensing algorithms.  Thus, it is desired to have a generalized hardware platform, which can be easily adjusted to implement different compressive spectrum sensing algorithms with different types of measurement matrices and recovery algorithms.

\section{Compressive Sensing enabled Large-Scale Wireless Sensor Networks}\label{IV}
WSNs provide the ability to monitor diverse physical characteristics of the real world, such as sound, temperature, and humidity, by incorporating information and communication technologies (ICT), which are especially important to various IoT applications. In the typical setup of WSNs, a large number of inexpensive and maybe individually unreliable sensor nodes with limited energy storage and low computational capability are distributed in the smart environment to perform a variety of data processing tasks, such as sensing, data collection, classification, modeling, and tracking. Cyber-physical systems  (CPSs) merge wireless communication technologies and environment dynamics for efficient data acquisition and smart environments control. Typically, a CPS consists of a large number of sensor nodes and actuator nodes, which monitor and control a physical plant, respectively, by transmitting data to an elaboration node, named local controller (LC) or FC.

Traditional environment information monitoring approaches take sensing samples  at a predefined speed uniformly at power-constrained sensor nodes and then report the data to a LC/FC, which is normally powerful and is capable of handling complex computations. The data transmitted  to the LC/FC usually  have redundancies, which can be exploited to reduce power consumption for data transmission. A common and efficient method is to compress at each individual sensor node  and then transmit. However, data compression introduces additional power consumption for individual sensor node although the power consumption on data transmission is reduced. Furthermore, this approach is unsuitable for real-time applications owing to the high latency in data collection and the high computational complexity to execute a compression algorithm at the power-constrained sensor node.

It is noted that most natural signals can be transformed to a sparse domain, such  as discrete cosine domain or wavelet domain, where a small number of the coefficients can represent most of the power of the signals that used to be represented by a large number of samples in their original domains. In fact, the data collected at each sensor node show a sparsity property in the discrete cosine domain or the wavelet domain due to the temporal correlation. Inspired by this, CS can be applied at each sensor node to collect the compressed measurements directly and then send to the LC/FC or the neighbour nodes. As a result, fewer measurements are sampled and transmitted and the corresponding power consumption is reduced significantly.

Power consumption of sensor nodes mainly comes from  sensing,  data processing, and communications with the LC/FC. In a large-scale WSN, the sensor nodes with low power level would wish to take samples at a lower speed or even turn themselves into the  sleep mode in order to extend  lifetime. As the signal received at each sensor node shows temporal correlation and neighboring nodes show spatial correlation, the joint sparsity can be exploited to recover signals from all sensor nodes even though samples from part of the participating sensor nodes are missed. The active sensor nodes can be pre-selected according to their power levels, therefore, with invoking of CS in WSNs,  lifetime of sensor nodes and the whole network can be extended.

%In order to get the full information of the network with less number of active sensor nodes,

The existing work on CS-enabled WSNs mainly falls into the aforementioned two categories: data gathering and active node selection, which will be introduced subsequently.

\subsection{Data Gathering}
In the traditional setting up of data gathering, there are a large number of sensor nodes  deployed in a WSN to collect monitoring data. Each sensor node generates a reading periodically and then sends  it to the LC/FC, which is normally powerful and capable of conducting complex computation. As the sensor nodes are generally limited in computation  and energy storage, the data gathering process in WSNs should be energy efficient with low overhead, which becomes very challenging in IoT scenarios where a huge number of sensor nodes are deployed. Taking the temperature monitoring as an example, the adjacent sensors will generate the similar readings as the temperatures of nearby locations are similar. Furthermore, for each sensor node, the readings from adjacent  snapshots are close to each other. The above two important observations indicate the temporal-spatial correlations among temperature readings, which enables the application of CS to reduce the network overhead as well as to extend the network lifetime. Moreover, such a joint sparsity is smaller than the aggregate over the individual signal sparsity, which results in a further reduction in the number of required measurements to exactly recover the original signals.

Instead of applying compression on the  data after it is sampled and buffered, each sensor node collects the compressed measurements directly by projecting the signal to its sparse domain.  At  each individual sensor node, one can naively obtain separate measurements of its signal and then recover the signal for each sensor separately at the LC/FC by utilizing the intra-signal correlation. Moreover, it is also possible to obtain compressed measurements that each of them is a combination of all signals from the cooperative sensor nodes in a WSN. Subsequently,  signals can be recovered simultaneously by exploiting both the inter-signal and intra-signal correlations at the LC/FC.

\subsubsection{Measurement Matrix Design}
When adopting CS techniques for data gathering  in WSNs, sampling at uniformly distributed random moments satisfies the RIP if the sparse basis $\Psi$ is orthogonal. For an arbitrary sensor node $i$, the $P \times N$ measurement matrix  can be a spike one that only has $P$ number of non-zero items, as shown by
\begin{align}\label{sensing_matrix}
\Phi_i  = \left[ {\begin{array}{*{20}{c}}
0&1&0&0& \ldots &0\\
0&0&0&1& \ldots &0\\
{}&{}&{}&{}& \ddots &{}\\
0&0&0&0& \ldots &1
\end{array}} \right].
\end{align}
Sensor node will take a sample at the moment when the corresponding item in $\Phi_i$ is one.

However, random sampling is not proper for practical WSNs since two samples may be too close to each other, which becomes very challenging for the cheap sensor nodes. In order to solve this issue, Chen and Wassell~\cite{Chen:IET2012} have proposed a  random sampling scheme by utilizing the temporal correlation of signals received at a sensor node. In the proposed scheme, the sensor node will send a pseudorandom generator seed to the FC and then send out the   samples that are obtained at an affordable highest rate until a sampling rate indicator (SRI) is received from the FC. Here, the SRI is decided based on the recovery accuracy calculated at the FC. Once the recovery accuracy goes out the required range, the sensor node will gradually increase its sampling rate  until the recovery error becomes acceptable. By adopting such a scheme, the  sensor node can adjust its sampling rate adaptively without the knowledge of the sparsity level. In order to further reduce the sampling rate at sensor nodes, spatial correlation is exploited in combination with  temporal correlation. Therefore, joint sparsity property can be exploited at the FC to reduce the number of required measurements.

More recently, a big data enabled WSNs framework has been proposed in~\cite{Kong:2016}, which invokes CS for data completion with a minimal number of samples. The proposed data collection framework consists of two core components: i) at the cloud, an online learning component predicts the minimal amount of data to be collected  to reduce the amount of data for transmission, and these data are considered as the principal data and their amount is constrained by CS; ii) at each individual node, a local control component tunes the collection strategy according to the dynamics and unexpected environment variation. Combining these two components, this framework can reduce power consumption and guarantee data quality simultaneously.

\subsubsection{Abnormal Sensor Detection}
\begin{figure*}[!t]
\centering
\includegraphics[width=5.8in]{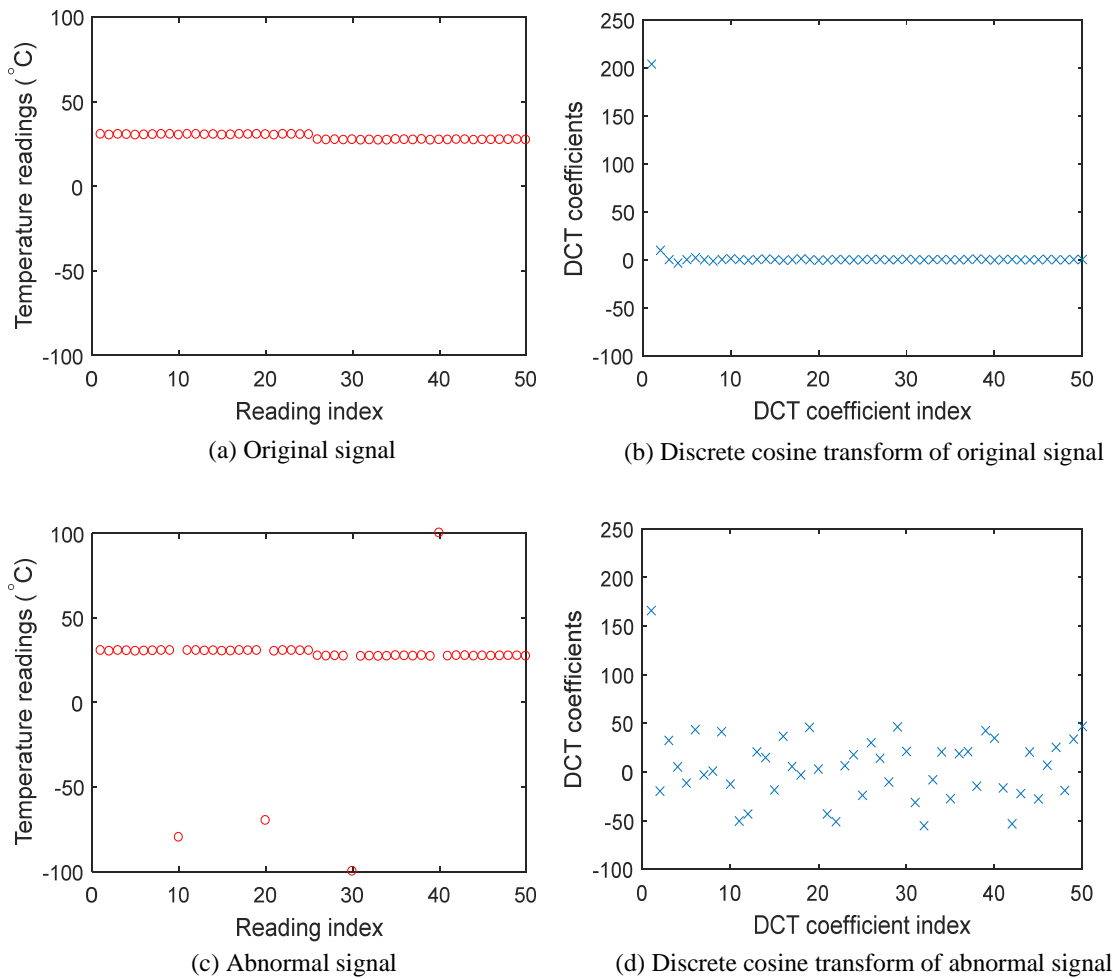}
\caption{Effect of abnormal readings on the sparsity level of temperature readings in discrete cosine domain.}
\label{abnormal}
\end{figure*}

In the CS-enabled data gathering processing, abnormal sensor readings may still lead to severe degradation on signal recovery at the LC/FC even if CS shows  robustness to abnormal sensor readings as it does not rely on the statistical distribution of data to be preserved during runtime.  This is because that the abnormal readings will damage the sparsity property of signals, as shown in  Fig.~\ref{abnormal}. Therefore, it is  critical to find out those abnormal sensors to guarantee the security of WSNs and make it abnormality-free.

Generally, abnormal readings are caused by either internal errors or external events according to their specific patterns. Abnormal readings due to internal errors fail to represent the sensed physical data, thus they should be removed at least. But the abnormal readings caused by the external events should be preserved as they reflect the actual scenarios of WSNs.

Inspired by recovering data from over-complete dictionary, an abnormal detection mechanism has been proposed to enhance the compressibility of the signals. First, the abnormal values are detected with the help of recovering signal from an over-complete dictionary. Second, the failing sensor nodes are categorized into different types according to their patterns. Thirdly, the failing nodes caused by the internal errors are removed and then the data recovery is carried out to obtain the ordinal data.

\subsection{Active Node Selection}
\begin{figure*}[!t]
\centering
\includegraphics[width=4.1in]{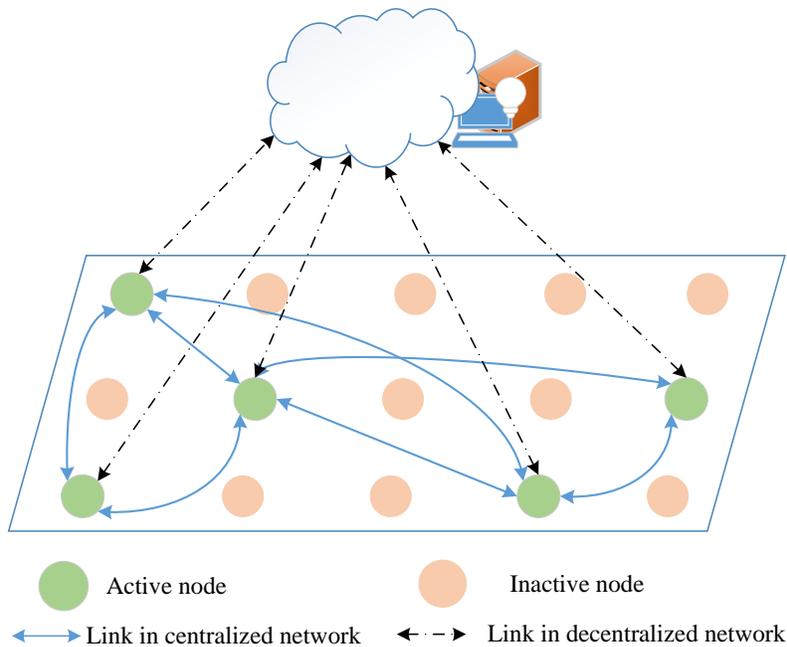}
\caption{Node selection in compressive sensing based wireless sensor networks.}
\label{node_selection}
\end{figure*}

In large-scale WSNs, the events are relatively sparse in comparison with the number of sensor nodes. Due to the power constraint, it is unnecessary to  activate all sensor nodes at all the time. By utilizing the sparsity property constructed by the spatial correlation, the number of active sensor nodes in each time slot can be significantly reduced without scarifying  performance. Taking the smart monitoring system as an example as shown in Fig.~\ref{node_selection}, the number of source nodes is $N$, and there are $K$ $\left(K\ll N\right)$ sparse events that are generated by the $N$ source nodes. By invoking CS, only $M$ $\left(M\le N\right)$ active sensor nodes are required to capture the $K$ sparse events.

\subsubsection{Centralized Node Selection}
A centralized node selection approach has been proposed in~\cite{Zhijin:TCOM2016}  by applying CS and matrix completion at the LC/FC with the purpose of optimizing the network throughput and extending the lifetime of sensors. As a node is either active or sleeping, the state index for a node  becomes binary, i.e., $X \in \left\{ {0,1} \right\}$. While conventional node selection in WSNs only exploits the spatial correlation of sensor nodes, Chen  and Wassell~\cite{Chen:2016TVT} have exploited the temporal correlation by using the support of the data reconstructed in the previous recovery period to select the active nodes. Specifically, the FC performs an optimized node selection, which is formulated as the design of a specialized measurement matrix, where the sensing matrix, $\Phi$, consists of selected rows of an identity matrix as shown in~(\ref{sensing_matrix}).

The sensing costs of taking  samples from different sensor nodes are assumed to be equal in  most of the node selection approaches. However,  in WSNs with power-constrained sensor nodes, this assumption does not hold due to the different physical conditions at different sensor nodes. For example, it is preferred to activate sensor nodes with adequate energy rather than those almost running out of energy to extend the lifetime of WSNs. Therefore, a cost-aware node selection approach has been proposed in~\cite{Chen:TSP2016} to minimize the sampling cost of the whole WSNs with constraints on the reconstruction accuracy.

\subsubsection{Decentralized Node Selection}
Different from the DCS, which normally conducts signal recovery at the FC by utilizing the data collected from the distributed sensing sources via exploiting the joint sparsity, decentralized CS-enabled WSNs approach aims to achieve in-network processing for node selection. A decentralized approach has been proposed in~\cite{Ling:TSP2010} to perform node selection by allowing each active sensor node to monitor and  recover its local data only, by collaborating with its neighboring active sensor nodes through one-hop communication, and  iteratively improving the local estimation until reaching the global optimum. It should be noted that an active sensor node  not only optimizes  for itself, but also  for its inactive neighbors. Moreover, in order to extend the network lifetime, an in-network CS framework has been developed in~\cite{Caione:TII2012} by enabling each sensor node to make an autonomous decision  on the data compression and forward strategy with the purpose of minimizing the number of packets to be transmitted.

Generally speaking, the drawbacks for distributed node selection approach come from the following two aspects: i) the optimized node selection requires an iterative process, which may require a long period; ii) the flexibility  to vary the number of active sensor nodes is limited, especially according to the dynamic sparsity levels  or the channel conditions, which could be time-varying. While for the centralized approach, extra bandwidth resource and power consumption are required to coordinate active sensor nodes.

\subsection{Potential Research}
Even though extensive research has been carried out to investigate the application of CS in WSNs, most of them have focused on  reducing power consumption at sensor nodes and extending the network lifetime. However, in large-scale WSNs for different IoT applications, big data should be exploited to enhance CS recovery accuracy in addition to further reduce the power consumption.

\subsubsection{Machine Learning aided Adaptive Measurement Matrix Design}
The core concept for sparse representation is the same even though different applications bring different constraints. Therefore, it is straightforward to ask if there is a general framework for the sparse representation of the data in different 5G and IoT applications for urban scenarios. In order to take the minimal number of samples from the set of sensor nodes with the best capability, i.e., highest power levels, the measurement matrix should be properly designed. It has been demonstrated  that machine learning can be an efficient tool to aid the measurement matrix design so that the lifetime of the whole network as well each individual sensor node can be extended to the most. Furthermore, it should be noted that when designing a measurement matrix, the possibility of implementation in real network is one of the most critical factors to be considered. We believe that extensive research work in this direction is highly desired.

\subsubsection{Data Privacy in CS-enabled IoT Networks}
The sensing data collected from a variety of featured sensors in IoT networks, our daily activities, surroundings, and even the physical information, can be recorded and analyzed, which at the same time greatly intensifies the risk of privacy exposure. There are few effective privacy-preserving mechanisms in mobile sensing systems. In order to provide privacy protection,  privacy-preserving noise has been proposed to be added to the original data  to guarantee effective privacy. The added noise will dominate the data when the original sparse data points are zero or near zero, thus reducing the data sparsity. However, CS aims to achieve kind of efficiency for sparse data processing. Then it comes into a conflict countable between privacy and efficiency during big data processing in CS-enabled IoT networks.  Therefore, extensive research work is expected in this area.

\section{Channel Acquisition and Precoding in Massive MIMO}\label{V}
To satisfy high data rate requirements caused by increasing mobile applications, a lot of efforts have been made to improve transmission SE. An effective way is to exploit the spatial degrees of freedom (DoF) provided by large-scale antennas at the transmitter and the receiver to form massive MIMO systems \cite{marzetta2010noncooperative}. It has been shown in \cite{marzetta2010noncooperative} that the spatial resolution of a large-scale antenna array will be very high and the channels corresponding to different users are approximately orthogonal  when the number of the antennas at the BS are very large. Consequently, linear processing is good enough to make the system performance to approach  optimum  if the CSI is known at the BS.

Accurate CSI at the BS is essential for massive MIMO to obtain the above advantage. Due to the large channel dimension, downlink CSI acquisition in massive MIMO systems sometimes becomes challenging even if uplink CSI estimation is relatively simple. In time-division duplexing (TDD) systems, the downlink CSI can be easily obtained by exploiting channel reciprocity. However, most of the practical deployed systems mainly employ {frequency-division duplex} (FDD), where the channel reciprocity does not hold any more. In this situation, the downlink channel has to be estimated directly and then fed back to the BS, which will result in the extremely high overhead.

To address CSI estimation and feedback issue in FDD systems, sparsity of massive MIMO channels must be exploited. Some CS-enabled CSI acquisition methods for FDD massive MIMO have been proposed, where the correlation in massive MIMO channels has been successfully exploited to reduce the amount of training symbols and feedback overhead.

In this section, we focus on the CS-enabled channel acquisition and its related applications. We will first introduce the channel sparsity feature and then discuss  channel estimation and feedback, and precoding and detection subsequently. It should be noted that mmWave communications are often used with massive MIMO techniques since short wavelength here makes it very easy to pack a large number of antennas in a small area. The channel acquisition and precoding based on CS in mmWave massive MIMO will be  also included in the following discussion even if mmWave channels are slightly different from the traditional wireless channels.

\subsection{Sparsity of Channels}

In the channel acquisition and precoding schemes based on CS, the key idea is to use the channel sparsity. Although the channel sparsity in massive MIMO generally exists in the time domain, the frequency domain, and the spatial domain, we mainly focus on the spatial domain channel sparsity in this section.

In conventional MIMO systems, a rich-scattering multipath channel model is often assumed so that the channel coefficients can be modelled as independent random variables. However, this assumption is not true any more in massive MIMO systems. It has been shown that the massive MIMO channel is spatial correlated and has a sparse structure in the spatial domain. This correlation and sparsity is due to two reasons: the exploitation of high radio frequency and the deployment of large-scale antenna arrays in future wireless communications. In high frequency band, the channels have fewer propagation paths while   more transmit and receive antennas will make the distinguished paths to be much fewer than the number of channel coefficients, which makes the rich scatters to become limited or sparse.

\begin{figure*}[!t]
  \centering
  \includegraphics[width=6.8in]{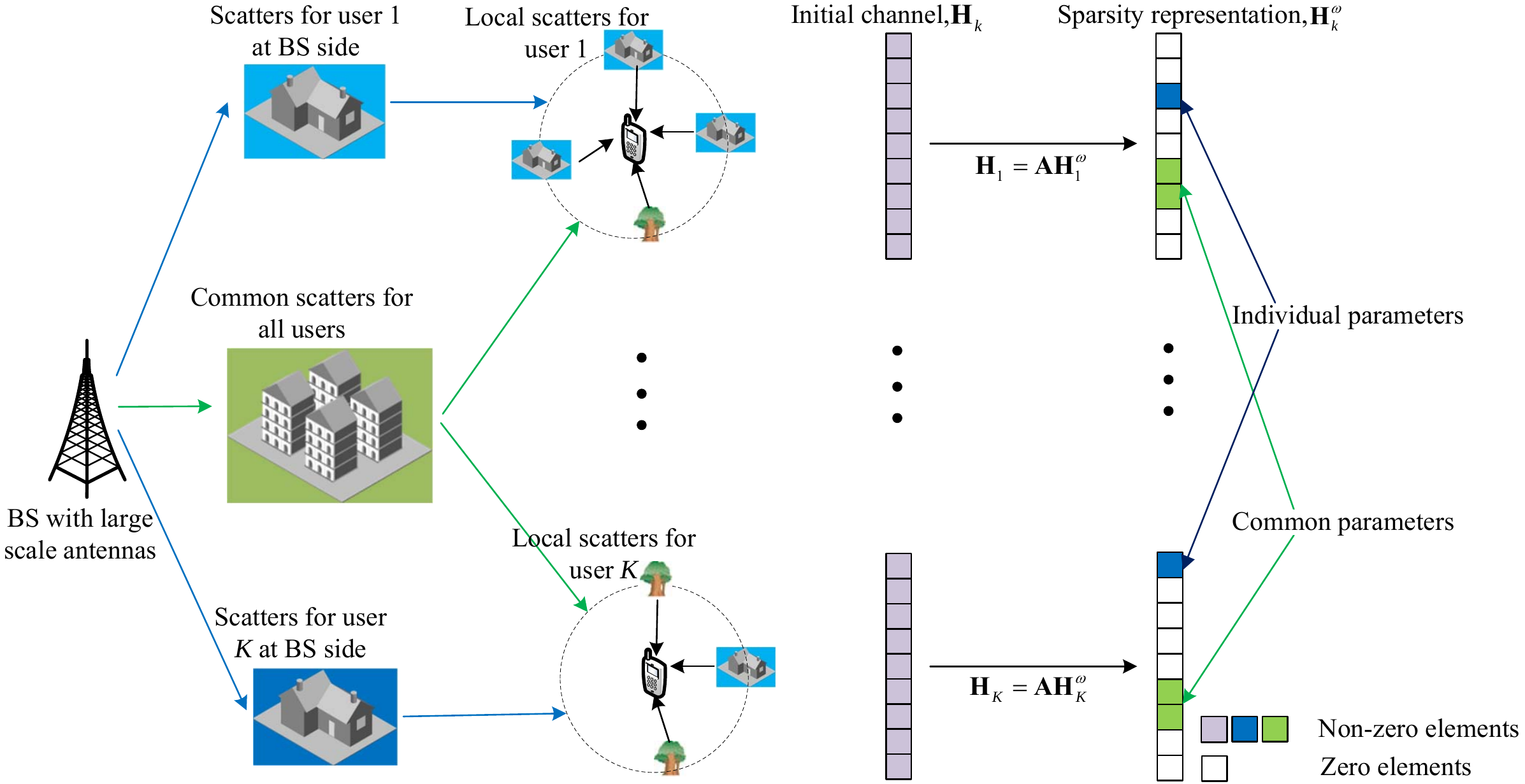}
\caption{Channel model with limited scatters, where $\mathbf{A}$ is the spatial correlation matrix.}
\label{channel model}
\end{figure*}

As shown in Fig.~\ref{channel model}, a classical channel model with limited scatterers at the BS is often used in the literature~\cite{liu2017closed}. In this model, different user channels have a partially common sparsity support due to the shared scatterers and an independent sparsity support caused by the individual scatterers in the propagation environments. By using this sparsity structure, many CS-enabled channel acquisition and precoding schemes have been proposed, as we will discuss in the following.

\subsection{Compressive Pilot Design}
In order to obtain a good channel estimation, the length of the orthogonal training sequence must be at least same as the number of transmit antenna elements. Due to a huge number of antennas at the BS in massive MIMO systems, the downlink pilots occupy a high proportion of the resources. Consequently, the traditional pilot design is not applicable here. It is necessary to design specific pilots to reduce the training overhead in FDD massive MIMO systems. It has been shown that channel spatial correlation or sparsity can be used to shrink the original channel to an effective one with a much lower dimension so that low-overhead training is enough in massive MIMO systems. Based on this principle, CS-enabled pilot design schemes have been proposed in~\cite{Lau2016ClosedLoop}.

When exploiting the CS theory to acquire the CSI in the correlated FDD MIMO systems, how much training should be sent is a very important question.  Once the amount of the training is determined, the training symbols can be designed by using of the channel common or/and individual sparsity in the correlated massive MIMO systems. Besides using the common support, a new pilot structure with joint common and dedicated support has been proposed in \cite{Lau2016ClosedLoop}, where the common one is used to estimate the common channel parameters of the related users in multi-user massive MIMO systems while the dedicated one is used to estimate the individual channel parameters of each user.

{Besides the general massive MIMO systems, a compressed hierarchical multiresolution codebook has also been designed specially for mmWave massive MIMO systems to construct training beamforming vectors \cite{alkhateeb2014channel}. Based on this idea, a lot of different hierarchical multiresolution codebooks have been proposed from different angles. For example, in \cite{marzi2016compressive}, a compressive beacon codebook with  different sets of pseudorandom phases has been designed. In \cite{song2017common}, a common codebook satisfying the conflicting design requirements as well as validating practical mmWave systems has been proposed through utilizing the strong directivity of mmWave channels. In \cite{wang2017low}, a multi-resolution uniform-weighting based codebook,  with similar to an normalized DFT matrix, has been proposed to reduce the implementation complexity, where the estimation of the angle of arrival (AoA) and angle of departure (AoD) will have a much lower training overhead. Therefore, the codebook based beamforming training procedure can achieve a good balance between complexity and high performance for the practical systems.}

\subsection{Compressive Channel Estimation and Feedback}
CS can be used to reduce the cost of the channel estimation and feedback by exploiting the channel sparsity. The existing CS based channel estimation and feedback schemes can be divided into the following three categories according to exploiting  sparsity in different domains.

\subsubsection{With Time Domain Sparsity}

\begin{figure*}[!t]
  \centering
  \includegraphics[width=7in]{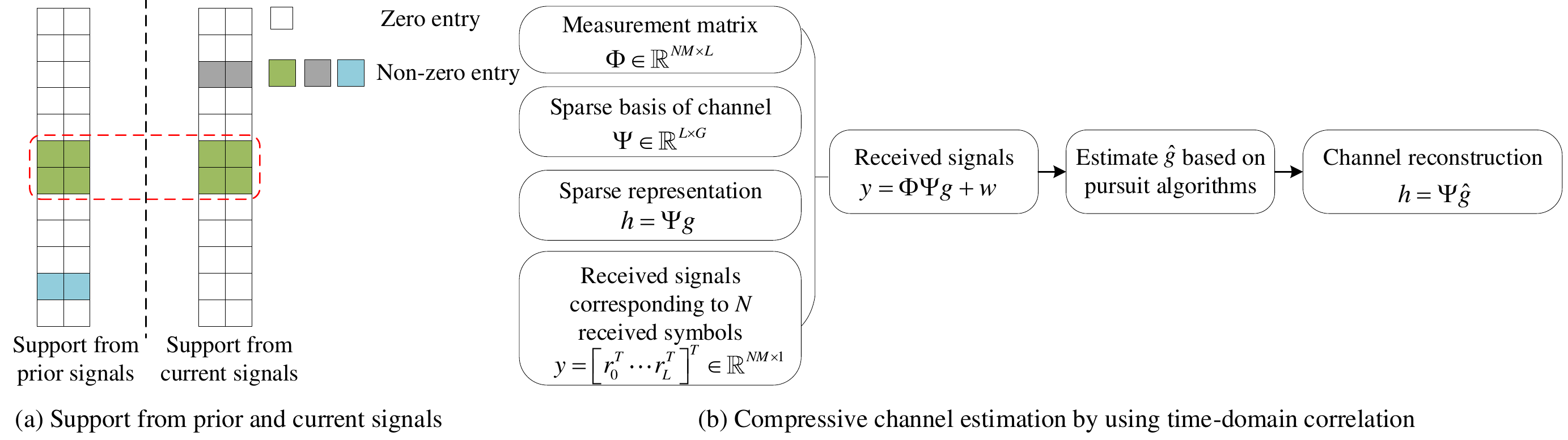}
  \vspace*{-0.3cm}
  \caption{Illustration of compressive channel estimation with time-domain correlation signal support.}\label{Time-correlation}
\end{figure*}

It has been shown that the channel is slowly-varying in various applications so that the prior channel estimation result can be still used to reconstruct the subsequent channels, as illustrated in Fig.~\ref{Time-correlation}. By using the structure in the figure, several CS algorithms have been proposed in \cite{shen2015compressive,han2017compressed,rao2015compressive} to recover massive MIMO channels.

Exploiting the common time-domain support, a CS-enabled differential CSI estimation and feedback scheme has been proposed in \cite{shen2015compressive}. The scheme in \cite{han2017compressed} further combines  LS and CS techniques to improve  estimation performance. In this scheme, the LS and CS techniques can be used to estimate a dense vector obtained by projecting into the previous support and a sparse vector obtained by projecting into the null space of the previous support, respectively. Since the current channel vector has only a small number of non-zero elements outside the support of the previous channel, the proposed scheme can reduce the pilot overhead and improve the tracking performance of the channel subspace. Since the quality of the prior support information will affect the estimation accuracy, a greedy pursuit-based approach with the prior support information and its quality information has been developed in \cite{rao2015compressive}, where the prior support information is adaptively exploited based on its quality information to further improve the channel estimation performance.

Besides  channel estimation,  several channel feedback schemes have been also proposed in \cite{shen2015compressive} and \cite{liu2017closed}. To exploit the common support, a CS-enabled differential CSI feedback scheme has been developed in \cite{shen2015compressive} by using the temporal correlation of MIMO channels. The proposed scheme can reduce the feedback overhead by about 20$\%$ compared with the direct CS-enabled channel feedback. In \cite{liu2017closed}, a robust closed-loop pilot and channel state information at the transmitter (CSIT)  feedback resource adaptation framework has been developed by using the temporal correlation of multiuser massive MIMO channels. In this framework, the  pilot and feedback resources can be adaptively adjusted for successful CSIT recovery under unknown and time-varying channel sparsity levels.

\subsubsection{With Spatial Domain Sparsity}
In practice, the antenna spacing in massive MIMO is usually set to be half-wave length to keep the array aperture compact. Furthermore, the BS with a large-scale antenna array is generally deployed at the top of high buildings such that there are only limited local scatters~\cite{Lu14An}. In this case, the large-scale MIMO channels exhibit strong angular-domain sparsity or spatial sparsity. This channel sparsity property can be used to reduce the channel estimation and feedback overhead in FDD massive MIMO systems.

In \cite{Tseng2016}, a block optimization algorithm is employed to extract the common angular support information from the channel matrices. The extracted common support information is then used to form the weighted factors and design a weighted block optimization algorithm to estimate the channel matrix. In~\cite{sim2016compressed}, a spatial sparsity-based compression mechanism has been proposed to reduce the load of the channel feedback. In this mechanism,  random projection with unknown sparsity basis and direct compression based on known sparsity basis are used. Since the spatial sparsity will reduce channel rank, a dictionary learning method, which captures the communication environment and the antennas property, has been proposed  to obtain the compressed channel estimation.

After the compressed hierarchical multiresolution codebook is designed in \cite{alkhateeb2014channel,marzi2016compressive,song2017common,wang2017low}, the corresponding channel estimation schemes can be developed. In these schemes, the hierarchical multi-resolution
codebook can be capable of generating variable beam width radiation patterns to facilate the usage of robust adaptive multipath channel estimation algorithms. Meanwhile, the exploitation of  adaptive compressive sensing algorithm will also reduce implementation complexity and estimation error.

\subsubsection{With Spatial-Temporal Sparsity}
In the above schemes, the temporal-domain sparsity and the spatial-domain one are independently exploited. In practice, they can be jointly used to further reduce the cost of channel estimation and feedback.

In \cite{shen2016joint}, a structured-CS enabled differential joint channel training and feedback scheme has been proposed, where a {structured compressive sampling matching pursuit} (S-CoSaMP) algorithm uses the structured spatial-time sparsity of wireless MIMO channels to reduce the training and feedback overhead. In \cite{huang2016rate}, a Bayesian CS based feedback mechanism has been proposed for time-varying spatially and temporally correlated  vector auto-regression  wireless channels, where the feedback rate can be adaptively adjusted.

\subsubsection{With Spatial-Frequency Sparsity}
Besides the spatial-temporal sparsity, the spatial-frequency one can also be exploited to reduce the cost of channel estimation and feedback. In \cite{kuo2012compressive}, the sparsity in the spatial-frequency domain is first exploited and an adaptive CS-enabled feedback scheme is correspondingly proposed to reduce the feedback overhead. In this scheme, the feedback can be dynamically configured based on the channel conditions to improve the efficiency. Due to sharing  sparse common support for the adjacent subcarriers in orthogonal frequency division modulation (OFDM), an  {approximate message passing with the nearest neighbor sparsity pattern learning} (AMP-NNSPL) algorithm has been developed to adaptively learn the underlying structure to obtain better performance.

\subsection{Precoding and Detection}
The precoder design based on the estimated CSI is a very important problem in massive MIMO systems, especially in the mmWave wideband systems. Since wideband mmWave massive MIMO channel generally exhibits frequency selective fading, the precoder design based on the estimated CSI will become challenging. Generally, the sparse structure of mmWave massive MIMO channels in angle domain or beam space can be used to simplify the precoder design \cite{venugopal2017optimality,mirza2017hybrid}. In \cite{venugopal2017optimality}, compressive subspace estimation is used to get the full channel information and then design the precoder to maximize the system SE. In order to reduce the CSI acquisition cost and address the mismatch between  few radio frequency chains and many antennas, the hybrid precoding design with baseband and radio frequency precoders has been used in mmWave massive MIMO systems. However, this hybrid precoding will introduce performance loss. In order to mitigate performance loss, an iterative OMP has been utilized to refine the quality of hybrid precoders \cite{mirza2017hybrid}. Meanwhile, a limited feedback technique has also been proposed for hybrid precoding to reduce the feedback cost.

CS can be also  used in signal detection of massive MIMO with spatial modulation (SM). In massive SM-MIMO, the maximum likelihood (ML) detector has a prohibitively high complexity. Thanks to the structured sparsity of multiple SM signals, a low-complexity signal detector based on CS has been introduced to improve signal detection performance. In \cite{gao2016compressive}, a joint SM transmission scheme for user equipment and a structured CS-enabled multi-user detector for the BS have been developed, the proposed detector can reliably detect the resultant SM signals with low complexity by exploiting the intrinsical sparse features.

In the above, we mainly discuss CSI acquisition and precoding based on CS theory in FDD massive MIMO systems. In practice, the CS theory can be also applied to TDD massive MIMO systems. A channel estimation approach based on block-structured CS has been developed in~\cite{nan2015efficient}, where the common support in sparse channels and the channel reciprocity in TDD mode are used simultaneously so that the computational complexity and pilot overhead can be reduced significantly.

\subsection{Potential Research}
As we can see from the above discussion, CS has been successfully used in massive MIMO to improve the performance of channel estimation  and precoding. However, there are still many open topics before implementing CS in massive MIMO systems. We will discuss some of them in this section.

\subsubsection{Effect of Antenna Deployment}
Due to space limitation, large-scale antenna may be deployed at various  topologies, i.e., centralized or distributed. Since the different antenna topologies corresponding to different channel sparsity, the effect of antenna configuration on the performance of CS-enabled channel estimation is still open.

\subsubsection{Measure of Channel Sparsity}
As mentioned before, the channel sparsity is very important to CS-enabled channel estimation. In the current research works, various sparsity models have been assumed and exploited in channel acquisition. Many of these assumed sparsity models lack of verification by measurement results. It is desired that either the sparsity model or the CS-enabled approach can be confirmed by some measurement results under different propagation conditions.

\subsubsection{Channel Estimation and Feedback with Joint Support}
In the above channel estimation and feedback schemes, one or two of the three types of channel sparsity are used to improve the channel estimation performance and reduce the feedback overhead. If all three types of channel sparsity are used, can we further improve performance?  How much performance gain can be achieved if possible?

\subsubsection{Channel Estimation and Feedback with Unknown Support}
Channel sparsity directly affects CS-enabled channel estimation performance. However, it is still a challenging problem how to get the information on channel sparsity support. At the same time, it is also worth to study how  channel estimation and feedback schemes should be designed without channel sparsity support.

\section{Other  Compressive Sensing Applications}
In addition to the aforementioned applications, CS has been applied in various other areas in wireless communications. Some of them  are introduced in this section.

\subsection{Compressive Sensing Aided Localization}
In a multiple target localization network, the multiple target locations can be formulated as a sparse matrix in the discrete spatial domain. By exploiting the spatial sparsity, instead of recording all received signal strengths (RSSs) over the spatial grid to construct a radio map from targets, only far fewer number of RSS measurements need to be collected at runtime. Subsequently, the target locations can be recovered from the collected measurements through solving an $\ell_1$ minimization problem. As a result, the multiple target locations can be recovered accordingly.

\subsection{Compressive Sensing Aided Impulse Noise Cancellation}
In certain applications, such as OFDM, impulsive noise will degrade the system performance severely. OFDM signal is often processed in the frequency domain. Even if impulse noise only lasts a short period of time, it affects a wide frequency range. By regarding impulse noise as a sparse vector, CS technique has been exploited to mitigate such type of impulse noise~\cite{Ramirez:EUSIPCO:2015}.

\subsection{Compressive Sensing Aided Cloud Radio Access Networks}
Cloud radio access networks (C-RANs) has been proposed as a promising technology to support massive connectivity in 5G networks.  In C-RAN, the BSs are replaced by  remote radio heads (RRHs) and connected to a central processor via digital backhaul links. Thanks to the spatial and temporal variation of the mobile traffic, it is feasible to switch off some RRHs with guarantee on the quality of service in green C-RANs. More specifically, one RRH will be switched off only when all the coefficients in its beamformer are set to zeros. Such a group sparsity property inspires us to apply CS to active RRH selection in green C-RANs to minimize the network power consumption~\cite{shi:2014:cran,Dai:2004IEEEACCESS}. Additionally, in the uplink of C-RANs, the channel estimation from the active users to the RRHs is the key to achieve the spatial multiplexing gain. Generally, the number of active users is low in C-RANs, which makes it possible to apply CS to reduce the uplink training overhead for channel estimation. Moreover, the correlation among active users at different RRHs exhibit a joint sparsity property, which can further facilitate the active user detection and channel estimation in C-RANs~\cite{Xu:ICC2015}.

%\subsection{Impulse Noise Cancellation}
%In addition to the aforementioned applications, by utilizing the corresponding sparse property, CS has been widely applied in both wireless communications and other areas, such as image processing, video processing, and classification due to its beautiful theory framework. We believe that CS will boost the signal processing in both wireless communications and beyond.

\section{Concluding Remarks}
This article has provided a comprehensive overview of sparse representation  with applications in wireless communications. Specifically, after introducing the basic principles of CS, the common sparse domains in 5G and IoT networks have been identified. Subsequently, three CS-enabled networks, including wideband spectrum sensing in CRNs, data collection in IoT networks, and channel estimation and feedback in massive MIMO systems, have been discussed by exploiting different sparsity properties. From the above discussions, it has been concluded  that by invoking of CS, the SE and EE of 5G and IoT networks can be enhanced from different perspectives. Furthermore, potential research challenges have been identified to provide a guide for researchers interested in the sparse representation in 5G and IoT networks.

\section*{Acknowledgment}
Jiancun Fan is supported by the National Natural Science Foundation of China under Grant 61671367, the China Postdoctoral Science Foundations under Grant 2014M560780 and Grant 2015T81031. Yue Gao is supported by funding from Physical Sciences Research Council (EPSRC) in the U.K. with Grant No. EP/R00711X/1.

\bibliographystyle{IEEEtran}
\bibliography{IEEEabrv,thesisbib}

\end{document}